# Pareto-Optimization Framework for Automated Network-on-Chip Design

Tzyy-Juin Kao[1] and Wolfgang Fink[1]

(1)Visual and Autonomous Exploration Systems Research Laboratory, University of Arizona, Department of Electrical and Computer Engineering, Tucson, AZ

## Abstract

With the advent of multi-core processors, network-on-chip design has been key in addressing network performances, such as bandwidth, power consumption, and communication delays when dealing with on-chip communication between the increasing number of processor cores. As the numbers of cores increase, network design becomes more complex. Therefore, there is a critical need in soliciting computer aid in determining network configurations that afford optimal performance given resources and design constraints. We propose a Pareto-optimization framework that explores the space of possible network configurations to determine optimal network latencies, power consumption, and the corresponding link allocations. For a given number of routers, average network latency and power consumption as example performance objectives can be displayed in form of Pareto-optimal fronts, thus not only offering a design tool, but also enabling trade-off studies.

## Introduction

Computational performance increases in single-core processors result in proportional increases in power consumption. Thus, the soaring power dissipation has become one of the performance bottlenecks. In contrast, multi-core processors achieve much higher computational power than single-core processors, harnessing the benefits of parallelism of computer tasks. When using several less-powerful small cores, with the ability to compute data simultaneously, the performance is boosted, at least in theory, proportionally to the number of cores with only a linear increase in power consumption.

Due to the rapid increase in numbers of processor cores, the data movement among all cores and memories forms another critical issue. Therefore, to keep improving the system performance, network-on-chip (NoC) design has gained considerable attention in recent years. Traditionally, linking all cores with a shared bus, through parallel electrical wires, causes serious congestion because every core uses the same link to transmit and receive data. Although there are many existing network topologies allocating resources with different pros and cons, most of them cannot be scaled efficiently to interconnect hundreds of cores on a chip. Therefore, many on-chip high-radix network topologies that feature low latency and better scalability have been proposed [1-3]. Implementing a well-designed network is crucial to meet the heavy-load bandwidth needs of multi-core processors.

In addition to the communication bandwidth, network performance in terms of latency and power consumption is critically affected by the chosen network topology. For example, by adding additional bypass links to a network it is possible to simultaneously reduce latency and power consumption when connecting two critical routers. Therefore, even if two network designs have an identical aggregate bandwidth, the latency and power performances may differ significantly, strongly depending on the respective link allocations. In light of the above, we propose a *Pareto-optimization framework (POF)* as an automated design tool for NoCs that explores different combinations of network configurations to determine link allocations to optimize network performance, i.e., to arrive at low-latency and power-efficient NoC architectures. This Pareto-optimization framework is an instantiation of a *Stochastic Optimization Framework (SOF)* devised by Fink in 2008 [4].

## Background

In the following we provide a definition and brief description of some of the technical terms used in this paper.

***Network-on-Chip (NoC)***

Figure 1 shows the schematic of a small 16-core NoC. Each core generates flits, i.e., smaller pieces of a packet, and transmits them to one designated router. Most flits usually have to pass through a few intermediate links and routers from a source node to a destination node because it is impractical and not cost-effective to implement a network that is fully connected, i.e., n(n-1) links, where n is the number of routers. Each router puts the receiving flits in buffers, waiting for the switch to direct them to the output to the destination node. Links are the connections between core-to-router and router-to-router, having parameters, such as latency and length, based on the physical inter-router distances on a chip. The activity of all components is recorded to calculate the total power consumption.

***Deterministic Routing Protocol***

The routing paths to send packets between any two routers are determined in advance. The most commonly used deterministic routing is *shortest path routing*: Packets follow the paths that have the shortest hop count without adapting to the current traffic load. The hop count is the performance unit to represent the total number of relay routers between any two nodes that a data packet must pass through. The benefits of this protocol are its simplicity and robustness.

***Pareto-Optimal Front***

Economist Vilfredo Pareto proposed a concept that for any allocation of resources there exists an optimal solution where no further improvement can be made without sacrificing one of the performance objectives in a multi-objective system. Because these performance objectives are usually conflicting, the Pareto-optimal front represents the optimal solution boundary after all performance evaluations, thus enabling trade-off studies [5].

***Stochastic Optimization Framework***

Figure 2 shows the functional schematic of a *Stochastic Optimization Framework* (SOF) [4]: A SOF efficiently samples the parameter space associated with a model, process, or system (1) by repeatedly running the model, process, or system in a forward fashion, and (2) by comparing the respective outcomes against a desired outcome, which results in a fitness measure. The goal of the SOF is to optimize this fitness by using multi-dimensional optimization algorithms, such as Simulated Annealing [6, 7], as the optimization engine to determine optimal parameter values. One way to determine optimal parameter values is to analytically invert models, processes, or systems, or to run them backwards. However, in many cases – as is the case with NoC design – this is analytically or practically infeasible due to the inherent complexity and high degree of non-linearity. A SOF overcomes this problem, by effectively "inverting" these models, processes, or systems to determine parameter values that, when applied, yield the desired outcomes, or approximate them as closely as possible (Figure 2).

## NoC Pareto-Optimization Framework Setup

A C++ program was developed building upon an open source network simulator *BookSim2.0* [8] to explore link allocations and resulting network performance (here: latency and power consumption) for any given number of routers on a chip. To quickly iterate the network simulations and evaluate the performance of each configuration, the simulation adopts *synthetic traffics (uniform random)*, instead of real application traffics. Compared to synthetic traffics, real application traffics provide more realistic results, but will take a longer time to evaluate. Furthermore, to efficiently obtain optimal results, instead of searching all combinations exhaustively, the program employs three optimization algorithms that are detailed below:

- Random Search (RS);
- Special Greedy (SG) as a deterministic optimization algorithm;
- Simulated Annealing (SA) [6, 7] as a stochastic optimization algorithm.

The program records all lowest possible network latencies in each power consumption interval, along with the corresponding NoC architectures, i.e., link allocations, for further analysis.

The overall flow of the proposed Pareto-optimization framework (Figure 3) is as follows:

1. Once the simulation starts, one of the three optimization algorithms creates an initial network configuration and feeds it to the BookSim2.0 simulator.
2. BookSim2.0 starts setting up objects of nodes, routers, and links based on this configuration and starts issuing packets from each node to another according to a pre-selected traffic pattern.
3. Those packets, generated according to the chosen traffic pattern, are routed from their source nodes to their destination nodes via the shortest paths that are calculated by the deterministic routing protocol.
4. When BookSim2.0 produces the simulation results, the optimization algorithm will create the next network configuration based on the current and previous performance objective values (i.e., latency and power consumption).
5. The optimization algorithm stores the best latencies for each power interval in a log file.
6. This loop (2–6) will keep running until a termination condition is met (i.e., desired performance objective values are reached), or a maximum number of iterations is exceeded.

Given the number of routers, the number of possible links can be derived from $n(n-1)/2$, where n is the number of routers. Then, the Pareto-optimization framework determines the link allocation to form the NoC. The link allocation is represented by a p-tuple, where p is the number of possible links. The elements of the p-tuple contain 0 and 1, representing absent and present links, respectively. The resulting number of combinations equals $2^{\frac{n(n-1)}{2}}$. For example, the link allocation of a 9-router network is a 36-tuple, resulting in about 68.7 billion combinations. Figure 4 shows the number of total network combinations as the number of routers increases. Therefore, facing this combinatorial explosion, an optimization algorithm to efficiently approximate the optimal result is critical.

The program employs three optimization algorithms:

a) **Total random:** The purpose of this random search is to give a baseline result for other optimization algorithms to compare against. It is simple to implement. In each loop, it generates a random link allocation, simulates this network configuration, and stores the results in a table if the network is stable and has a better latency than the best NoC encountered so far.

b) **Special greedy:** This deterministic algorithm starts from a fully-connected point-to-point network. All links are present and it is guaranteed to obtain the lowest latency of all possible link allocations. It is the optimal solution in terms of latency with the caveat of a high power consumption. Then, the following iterations explore neighboring networks (i.e., networks that have only one link fewer as the current network) one by one and move to the neighbor that has the minimum latency increase. The algorithm continues until there is no stable neighboring network and no further link can be removed. The greedy algorithm will yield a local minimum eventually because the next location is based on a local condition and the history of deterministic choices. However, it usually has many fewer steps than other stochastic or non-deterministic algorithms. For example, a 9-router network has 36 possible links and all are present at first. In the first iteration, the program simulates all 36 neighboring networks, which have one link removed. The following iterations simulate from 35, 34, ..., to 9 neighboring networks. No links can be removed when the number of links has reached 8, i.e., n-1. The total number of simulations for this algorithm is $(n^4 - 2n^3 - n^2 + 2n)/8$, where n is the number of routers.

c) **Simulated annealing [6, 7]:** This algorithm starts from a random link allocation and keeps moving to neighboring networks based on a stochastic condition until the temperature variable is cooled below a terminating point. The neighboring network is created by flipping one random location/link of the current p-tuple of link allocations (i.e., a random element changed from 1 to 0 or vice versa). Then, the decision of either ACCEPT or REJECT the new network is made by comparing the fitnesses of two networks. The fitness E is the weighted summation between performance objectives (here: latency and power). If the new network has a better fitness, the algorithm accepts it and moves on, but if it has a worse fitness, the algorithm will make a decision based on a Boltzmann-probability:

$$\text{random number } [0\ldots1] < \exp[-(E_{temp} - E_{best})/T]$$

where $E_{temp}$ is the new fitness, $E_{best}$ is the best fitness recorded so far, and T is the temperature variable. The temperature is decreasing after each iteration according to a cooling rate (λ): When the temperature is high, the algorithm has a high probability of accepting a worse network, and as the temperature decreases gradually, the probability of accepting a worse network also decreases. Then, only networks with better fitness will be accepted. Therefore, because of the feature of a Boltzmann-probability, simulated annealing can avoid getting stuck in local minima, thus approaching the global minimum.

In our simulations (see Results) we preset a start and end temperature as well as a cooling rate λ such that the maximum number of iterations is about one million. In addition, the fitness has tunable multi-objective weights to explore areas of design space the user is interested in (i.e., latency and power consumption in our case). For our simulations, the overall fitness of each network design is expressed as a weighted summation equation of latency and power consumption:

$$\text{Fitness } E = \text{weight} \times \text{latency} + (1 - \text{weight}) \times \text{power} \quad .$$

Because the simulated annealing algorithm iterates and converges to one optimal result (lowest fitness) eventually, based on its tunable multi-objective weights in the fitness, we sweep the weight parameter from 0.1 to 1 in increments of 0.1 to generate a Pareto-optimal front across a wide range of power consumption. Hence, when the weight is low (0.1 or 0.2), the algorithm explores the leftmost side of the Pareto-optimal front where the power is low, and when the weight increases, it moves to the right gradually where the latency is low. Programs with different weight settings can be executed in parallel on a cluster computer and can record the results simultaneously, generating a Pareto-optimal front at last. The weight cannot be 0, i.e., power only, because without taking latency into consideration, the simulated annealing gets into unstable networks with diverging latencies.

Because of the use of BookSim2.0, all resulting networks guarantee that any two routers can be connected through other routers (i.e., no orphans and no isolated groups). While the program keeps simulating different networks, two performance objectives, i.e., latency and power consumption, are monitored for Pareto-optimization analysis. Power consumptions in watt are rounded to the nearest integer to reduce the overall data that need to be recorded. Therefore, in each integer power interval, only the minimum latency is recorded.

BookSim2.0 requires to manually assign the latencies (in cycles) of all links since it cannot calculate the time delay for a signal to travel from one end to another of a link based on its length. Therefore, DSENT [9] is used in our simulations to calculate the minimum required latencies for different physical lengths of links in advance. Assuming all links on a chip only stretch into horizontal and vertical directions (not diagonal for better layout formatting) and all routers are distributed evenly across the chip (i.e., tiled architecture [10]), then each inter-router-distance (i.e., rectangular directions) can be derived from the die size and the number of routers. Whenever a link allocation is generated, the program calculates link latencies based on their inter-router-distances, and then creates a complete network configuration for BookSim2.0 to simulate.

All network devices are based on a 32 nm CMOS technology and assumed to operate at 5 GHz on a 21 x 21 $mm^2$ chip for performance analysis. Table 1 below shows the configurations on BookSim2.0. *Anynet* is one of the topology functions that reads the configuration file:

| **Variable:** | **Value:** |
|---|---|
| Topology | anynet |
| Routing function | min |
| Traffic | uniform |
| Sample period | 1000 |
| Injection rate | 0.1 |

*Min routing function* is the deterministic routing protocol that generates routing path tables based on the shortest hop counts between routers. *Uniform traffic* is the random synthetic traffic pattern. All packets are generated randomly based on the injection rate of each router and are sent to a random destination. *Sample period* is the cycle time of each simulation. In addition, to ensure the network is stable (i.e., generates converging results), every network is simulated at least four times. If one of the results is diverging, the simulator will discard it and run another time. *Injection rate* is the frequency of a new packet generated by each node.

## Results

Figure 5a shows the Pareto-optimal front between two performance objectives, i.e., latency as a function of power consumption, for a 16-router network scenario. First, a POF-designed NoC with lower latency but the same power compared to the well-known mesh network is found. A mesh network consists of short links only. In contrast, the POF-designed NoC contains some longer links, which reduce the number of relay routers between some routers. A lower number of relay routers results in a decrease in latency and power consumption for writing data into the router buffers. The POF is capable of finding the balance between the performance objectives. Second, simulated annealing produces better results than the other two algorithms. A fully-connected point-to-point (pt2pt) network is also plotted as a reference to show the lowest possible latency of all networks.

Figure 5b shows the Pareto-optimal front between two performance objectives, i.e., latency as a function of power consumption, for a 36-router network scenario. Because the number of links to form a 36-router network is between 35 to 630, the range of power consumption increases to a maximum of 350 watts (a fully-connected point-to-point network), which is an unaffordable cost for a multi-core processor. When the power increases to about 45 watts, the latency decreases from about 23 to 18 cycles. Hence, those networks (in the small figure on the left) are more advantageous network designs. When the power keeps increasing, the latency only decreases slightly (i.e., flat Pareto-optimal front), indicating that the rest of the designs are not significantly more power-efficient. On the other hand, the POF-designed NoC also outperforms the mesh NoC, and this gap has increased more compared to the 16-router network scenario above. In addition, simulated annealing still finds better networks than the greedy algorithm, which indicates the greedy algorithm may have gotten stuck in a local minimum as anticipated.

Figure 5c shows the Pareto-optimal front between two performance objectives, i.e., latency as a function of power consumption, for a 64-router network scenario. Compared the mesh network with the lowest-power POF-designed NoC, although mesh only requires 24 watts and the POF-designed NoC requires 44 watts, there is a huge latency tradeoff (from about 197 to 24 cycles). The result shows that the mesh network has reached its saturation point [11] to accommodate current traffic load, and is a poor design based on the Pareto-optimal front. Other networks that use hundreds or thousands of watts are also impractical designs due to the power consumption budget on a single chip. Fully-connected pt2pt network demands the highest power, more than 3000 watts, resulting from placing all 2016 links between any two routers.

The weights for the simulated annealing algorithm to generate the Pareto-optimal fronts for the 16-, 36-, and 64-router scenarios are shown in Table 2:

| **16-router scenario** | | **36-router scenario** | | **64-router scenario** | |
|---|---|---|---|---|---|
| Weight | Power Range | Weight | Power Range | Weight | Power Range |
| 0.1 | 7 | 0.2 | 19–21 | 0.1 | 44–45 |
| 0.4 | 8 | 0.5 | 22–25 | 0.2 | 46–48 |
| 0.6 | 9 | 0.7 | 26–29 | 0.5 | 49–52 |
| 0.7 | 10–11 | 0.8 | 30–39 | 0.7 | 53–59 |
| 0.8 | 12–13 | 0.9 | 40–73 | 0.8 | 60–71 |
| 0.9 | 14–17 | 1.0 | 74–188 | 0.9 | 72–424 |
| 1.0 | 18–25 | | | 1.0 | 425–746 |

The POF-designed network topologies (i.e., link allocations) are also recorded during each simulation, allowing for the actual display of their design. The lowest power NoCs found by simulated annealing are plotted in Figure 6: Black lines represent the links that a mesh network topology has in common, and blue lines represent the opposite to clearly display the differences from the mesh network. The figure shows the network topologies indicating the connections between routers, not the actual physical layout of the electric wires. Typically, electric wires are laid-out in horizontal and vertical directions on a chip distributed among multiple metal layers to avoid crosstalk, and wires in different metal layers are connected by vias (vertical interconnect access). For example, the longest link in Figure 6a, connecting router 5 and router 15, is 4 inter-router-distances long, i.e., 2 inter-router-distances in both horizontal and vertical directions. Because longer links consume higher power consumptions, simulated annealing starting from a random network still converges to a mesh-link network that uses many one-inter-router-distance links to connect neighboring routers. The results also show that the mesh topology has limited scalability because as the number of routers increases, one-inter-router-distance links are no longer efficient in terms of latency and power consumption, and are gradually replaced by longer links (Figure 6b, c). When packets are routed, long links lower the latency and reduce the usage of relay routers.

## Discussion

To appreciate, assess, and corroborate the feasibility and quality of the NoC designs found by the introduced Pareto-optimization framework the following has to be emphasized:

1. The BookSim2.0 simulator always guarantees that a network is stable before producing latency and power results [12]. A stable network means that all packets can be sent to the correct destinations within a given (finite) simulation time. Hence, no packet is lost or stuck in the network when following the given routing protocol.
2. The deterministic routing protocol, which always directs packets along the same path between any two routers, is adopted and works properly even for these unconventional and irregular networks that are created by the Pareto-optimization framework. The protocol calculates the minimal paths among routers in advance and follows them even if there are other possible paths. In doing so, any *livelock loop* (i.e., a packet is travelling around and never reaching the destination) is prevented and at least all path lengths are the shortest. If any *deadlock* occurs (i.e., some packets hold resources and also request other resources that are held by other packets; if their dependencies form a cycle and none of these packets can proceed, a deadlock is created permanently), the simulator will abort those unstable networks.

3. The BookSim2.0 simulator simulates a network at least four times to make sure a network can produce converging results under different random traffics. Therefore, the simulated network should be robust enough to allow smooth traffic flow without sudden congestion and uneven delay in certain paths, thus lending credibility to the latency results.

The Pareto-optimization framework uses *random synthetic traffic* to quickly evaluate the latency and power consumption among billions of network combinations. Although more sophisticated multi-core processor simulators exist, such as gem5 [13], they usually are computationally much more expensive than BookSim2.0, i.e., their incorporation in the Pareto-optimal framework, albeit feasible in principle, is computationally rather unrealistic unless they were amenable to parallelization. Therefore, it is advantageous to use a computationally cheaper simulator, such as BookSim2.0, at first to optimize NoC architectures much more rapidly in an iterative manner, and to subsequently benchmark the resulting optimal POF-designed NoC architectures with a sophisticated and comprehensive NoC simulator, such as gem5 (see section 6 below for justification of this procedure).

Given the quartic equation of the number of iterations $(n^4 - 2n^3 - n^2 + 2n)/8$, the special greedy algorithm has become difficult to complete for larger router numbers: For a 64-router scenario, it already requires 2 million iterations per simulation, i.e., twice the number as the simulated annealing algorithm, and for a 256-router scenario 532 million iterations, thus becoming quickly impractical to use. In contrast, the complexity of the simulated annealing is based on the user-defined temperature parameters and the cooling rate, i.e., independent of the number of routers. In addition, compared to special greedy, simulated annealing produces better results within an adjustable finite simulation time.

Until now, POF-designed PoCs tend to adopt more links to create networks with lower latency (i.e., fewer cycles) at the expense of power consumption. However, when restricting the number of links to be 112 for a 64 router network, i.e., the same as a 64 router (8 x 8) mesh network, we found so far no better but similarly performing POF-networks in our POF simulations, as shown in Figure 7. Mesh networks still have many other benefits that POF-networks do not have, such as that they can utilize many different routing protocols and exhibit high path diversity and fault-tolerance due to the link symmetry. POF-designed PoCs though show that there exist a lot more alternative networks to be considered while also allowing users to add restrictions based on their respective application needs.

## Full System Application Benchmarking

Although the introduced POF, when using BookSim2.0, has the ability of quickly evaluating billions of network combinations and generating a full Pareto-optimal front, these simulations are only conducted under random synthetic traffic as previously discussed. Therefore, to provide detailed evaluations of a system under real-world applications and to validate our POF simulation results, we further simulated and compared a standard mesh NoC architecture and the lowest-power 16-router POF-designed NoC architecture using the *full system cycle accurate gem5 simulator* [13].

The gem5 simulator consists of CPU models, a detailed cache/memory system, on-chip interconnection networks (including links, routers, routing protocols, and flow-control), and a variety of cache coherence protocols. In the full system mode, gem5 builds a system based on configuration input (e.g., our POF-designed NoC) and boots a Linux operating system on it – all in virtual space. Application benchmarks are then executed at runtime of the operating system. We select the *PARSEC benchmark suite* [14, 15] for our NoC benchmarking due to its emerging parallel applications, especially for multi-core processors. Table 3 below shows the system configuration for the 16-router system on gem5.

| | | | |
|---|---|---|---|
| **Core count:** | 16 and 64 | **Core frequency:** | 2GHz |
| **L1 I/D cache:** | 32KB/32KB | **L2 cache per core:** | 2MB |
| **Cache coherence:** | Decentralized directory based | **State machine:** | MOESI |
| **Link latency:** | Proportional to inter-router-distance | **Link bandwidth:** | 128 bits |

We implement the POF-designed NoC custom topology in gem5 to simulate/assess its average network latency while running the PARSEC benchmark suite. Power consumption is evaluated by the Orion2.0 power simulator [16] interfaced with gem5. Figure 8 shows the evaluations of both 16- and 64-router scenarios. The POF-designed NoC exhibits a similar latency reduction compared to mesh NoC when using gem5 instead of BookSim2.0. The traffic loads, generated by the random synthetic traffic and the chosen benchmarks, differ between the two simulators used (i.e., BookSim2.0 and gem5). Thus, the latency results differ quantitatively. However, they are consistent qualitatively. Furthermore, no deadlock occurs and the deterministic routing protocol is also working properly in the full system environment (i.e., gem5). The power consumption also shows similar results as the POF simulations. For the 16-router scenario, POF-designed NoCs have an average latency reduction of 3.02 cycles with a cost of 1.07 watts. For the 64-router scenario, POF-designed NoCs have an average latency reduction of 12.69 cycles with a cost of 24.95 watts. In addition, based on the previous POF simulation, POF-designed NoCs are expected to have higher saturation points, which enable the handling of higher traffic loads. Although mesh-NoCs have the advantage of a much simpler and regular architecture/link layout, POF-designed NoCs, which are selected from an enormous number of combinations beyond human/manual design abilities, prove to be superior in regards to latency and power consumption, and are implementable and controllable (protocol-wise).

## Conclusion

A Pareto-optimization simulation framework that automates NoC architecture design has been devised. When taking inter-router-distance into consideration, it is hard to find an efficient set of short and long links without the aid of a computer, especially in large-scale multi-core systems. Long links consume more power than short links, but they reduce the number of relay routers in a path between two routers, thereby decreasing the latency. Therefore, adding long links at opportune locations is critical. The framework is capable of iterating and exploring the tradeoffs between these two performance objectives (i.e., latency and power consumption) in form of a Pareto-optimal front. Among the three tested optimization algorithms, the simulated annealing algorithm is very efficient and has high flexibility because the weights for each performance objective can be fine-tuned to fulfill different application needs.

The Pareto-optimization framework shows encouraging results, paving the way towards fully automated NoC design. Additional design parameters, such as load balancing, adaptive routing protocol, and photonic links [17], can be considered and incorporated in future work to further enhance the scope and quality of automated NoC designs to meet the exploding need for multi-core systems.

## Competing Financial Interests

Authors T.-J. K. and W.F. may have a competing financial interest in the Pareto-Optimization Framework for Automated Network-on-Chip Design presented here as a provisional patent application has been filed on behalf of the University of Arizona.

## References

1. J. Kim, J. Balfour, and W. Dally, “Flattened butterfly topology for on-chip networks,” in 40th Annual IEEE/ACM International Symposium on Microarchitecture (MICRO 2007), Dec 2007, pp. 172–182.
2. Y. H. Kao, M. Yang, N. S. Artan, and H. J. Chao, “Cnoc: High-radix clos network- on-chip,” IEEE Transactions on Computer-Aided Design of Integrated Circuits and Systems, vol. 30, no. 12, pp. 1897–1910, Dec 2011.
3. B. Grot, J. Hestness, S. W. Keckler, and O. Mutlu, “Kilo-noc: A heterogeneous network-on-chip architecture for scalability and service guarantees,” in 2011 38th Annual International Symposium on Computer Architecture (ISCA), June 2011, pp. 401–412.
4. W. Fink (2008) *Stochastic Optimization Framework (SOF) for Computer-Optimized Design, Engineering, and Performance of Multi-Dimensional Systems and Processes*; SPIE Defense & Security Symposium, Orlando, Florida; Proc. SPIE, Vol. 6960, 69600N (2008); DOI:10.1117/12.784440 (invited paper)
5. A.V. Lotov and K. Miettinen, Visualizing the Pareto Frontier. Springer Berlin Heidelberg, 2008.
6. N. Metropolis, A.W. Rosenbluth, M.N. Rosenbluth, A.H. Teller, E. Teller, Equation of State Calculation by Fast Computing Machines, *J. of Chem. Phys.*, 21, 1087 – 1091, 1953.
7. S. Kirkpatrick, C.D. Gelat, M.P. Vecchi, Optimization by Simulated Annealing, *Science*, 220, 671 – 680, 1983.
8. N. Jiang, J. Balfour, D. U. Becker, B. Towles, W. J. Dally, G. Michelogiannakis, and J. Kim, “A detailed and flexible cycle-accurate network-on-chip simulator,” in 2013 IEEE International Symposium on Performance Analysis of Systems and Software (ISPASS), April 2013, pp. 86–96.
9. C. Sun, C.-H. Chen, G. Kurian, L. Wei, J. Miller, A. Agarwal, L.-S. Peh, and V. Stojanovic, “DSENT - a tool connecting emerging photonics with electronics for opto- electronic networks-on-chip modeling,” in Proc. 6th IEEE/ACM Int. Symp. Netw. Chip (NoCS), May 2012, pp. 201–210.

10. J. Balfour and W. J. Dally, “Design tradeoffs for tiled CMP on-chip networks,” in Proceedings of the 20th Annual International Conference on Supercomputing, ser. ICS’06. New York, NY, USA: ACM, 2006, pp. 187–198.
11. S. Park, T. Krishna, C.-H. Chen, B. Daya, A. Chandrakasan, and L.-S. Peh, “Approaching the theoretical limits of a mesh noc with a 16-node chip prototype in 45nm soi,” in Proceedings of the 49th Annual Design Automation Conference, ser. DAC ’12. ACM, 2012, pp. 398–405.
12. N. Jiang, G. Michelogiannakis, D. Becker, B. Towles, and W. J. Dally, “Booksim 2.0 user’s guide,” Stanford University, March 2010.
13. N. Binkert, B. Beckmann, G. Black, S. K. Reinhardt, A. Saidi, A. Basu, J. Hestness, D. R. Hower, T. Krishna, S. Sardashti, R. Sen, K. Sewell, M. Shoaib, N. Vaish, M. D. Hill, and D. A. Wood, “The gem5 simulator,” SIGARCH Comput. Archit. News, vol. 39, no. 2, pp. 1–7, Aug. 2011.
14. C. Bienia, S. Kumar, J. P. Singh, and K. Li, “The PARSEC Benchmark Suite: Characterization and Architectural Implications,” in Proceedings of the 17th International Conference on Parallel Architectures and Compilation Techniques, October 2008.
15. M. Gebhart, J. Hestness, E. Fatehi, P. Gratz, S. W. Keckler, “Running PARSEC 2.1 on M5,” The University of Texas at Austin, Department of Computer Science. Technical Report #TR-09-32. October 27, 2009.
16. Kahng, A. B., Li, B., Peh, L.-S., and Samadi, K. ORION 2.0: a fast and accurate NoC power and area model for early-stage design space exploration. In Proceedings of the Conference on Design, Automation and Test in Europe (2009), pp. 423–428
17. T. J. Kao and A. Louri, “Optical multilevel signaling for high bandwidth and power-efficient on-chip interconnects,” IEEE Photonics Technology Letters, vol. 27, no. 19, pp. 2051–2054, Oct 2015.

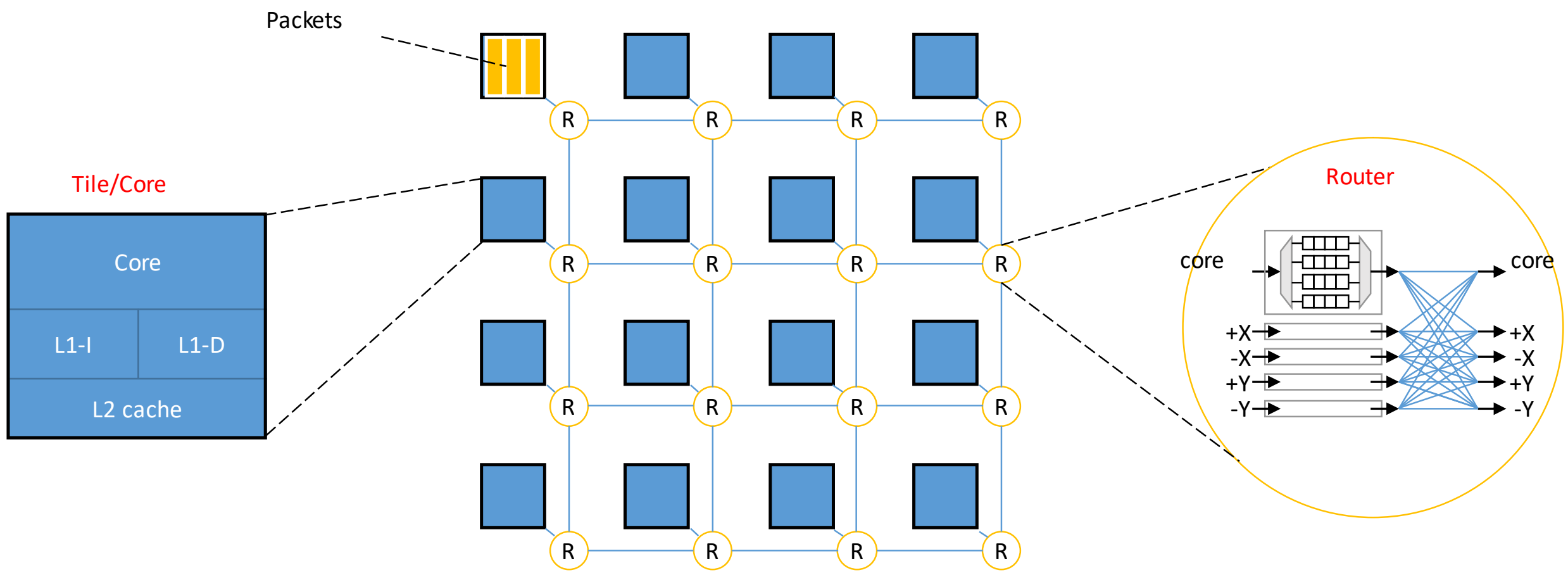

Figure 1. A typical example of a 16-core NoC.

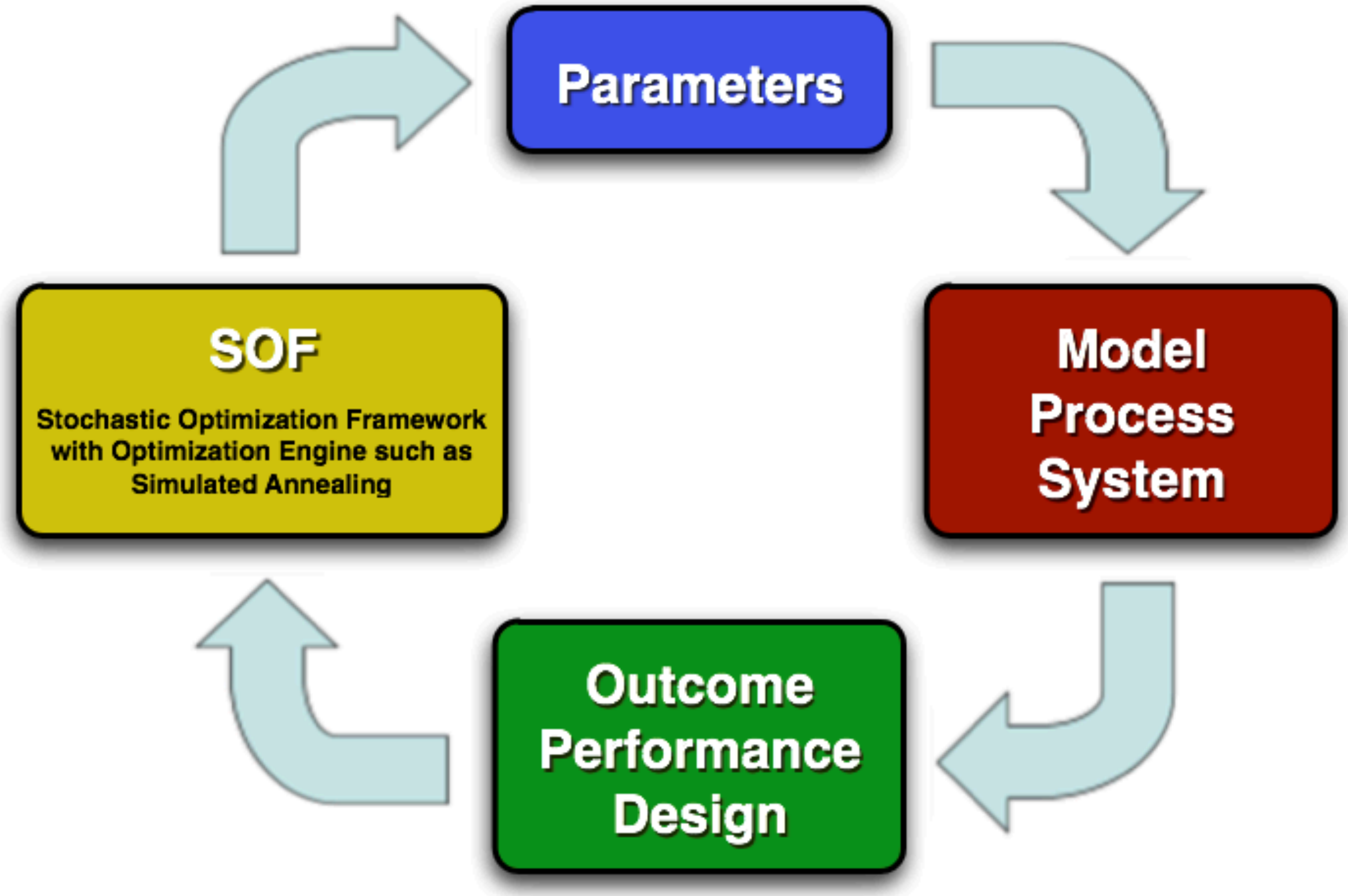

Figure 2. Functional schematic of a *Stochastic Optimization Framework* (SOF; after [4]).

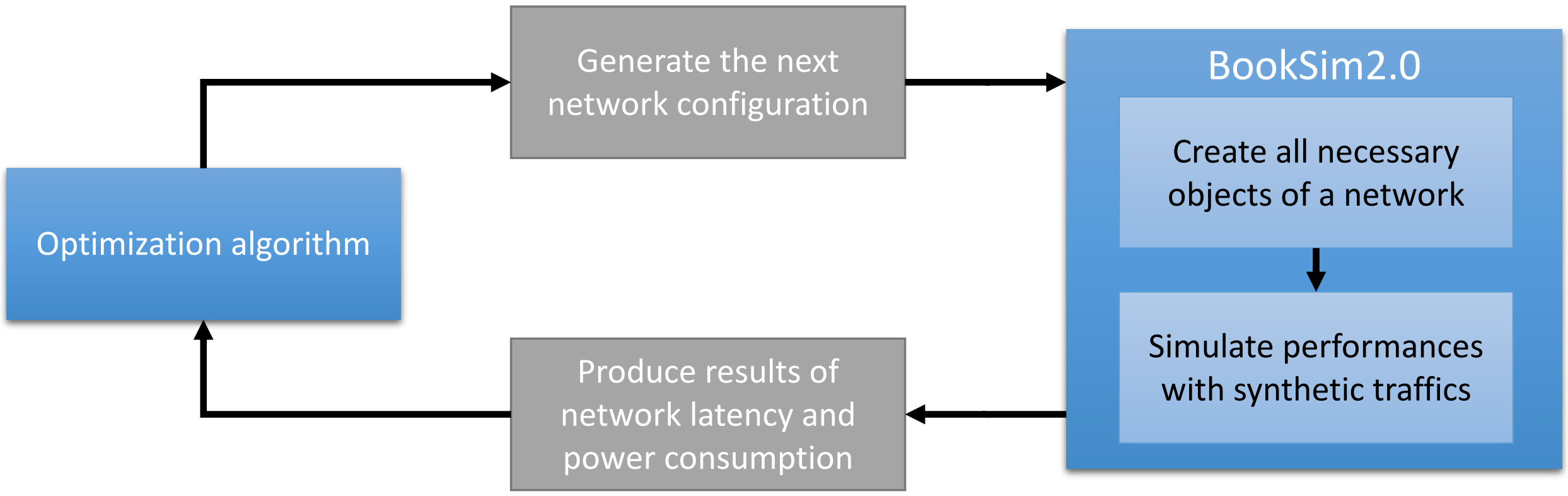

Figure 3. Schematic of the Pareto-optimization framework setup – an instantiation of a *Stochastic Optimization Framework* (SOF; after [4]).

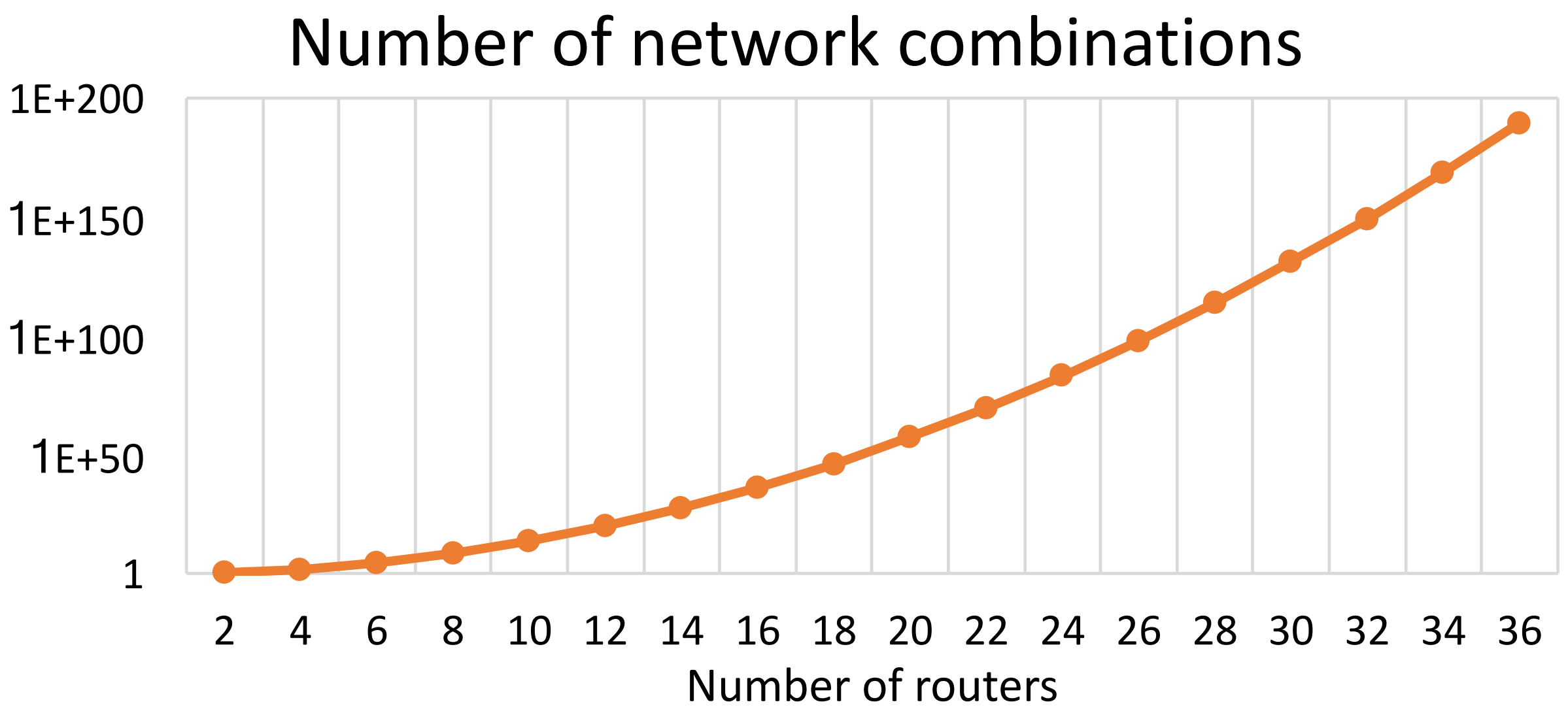

Figure 4. Total number of network combinations as a function of number of routers in a semi-logarithmic plot.

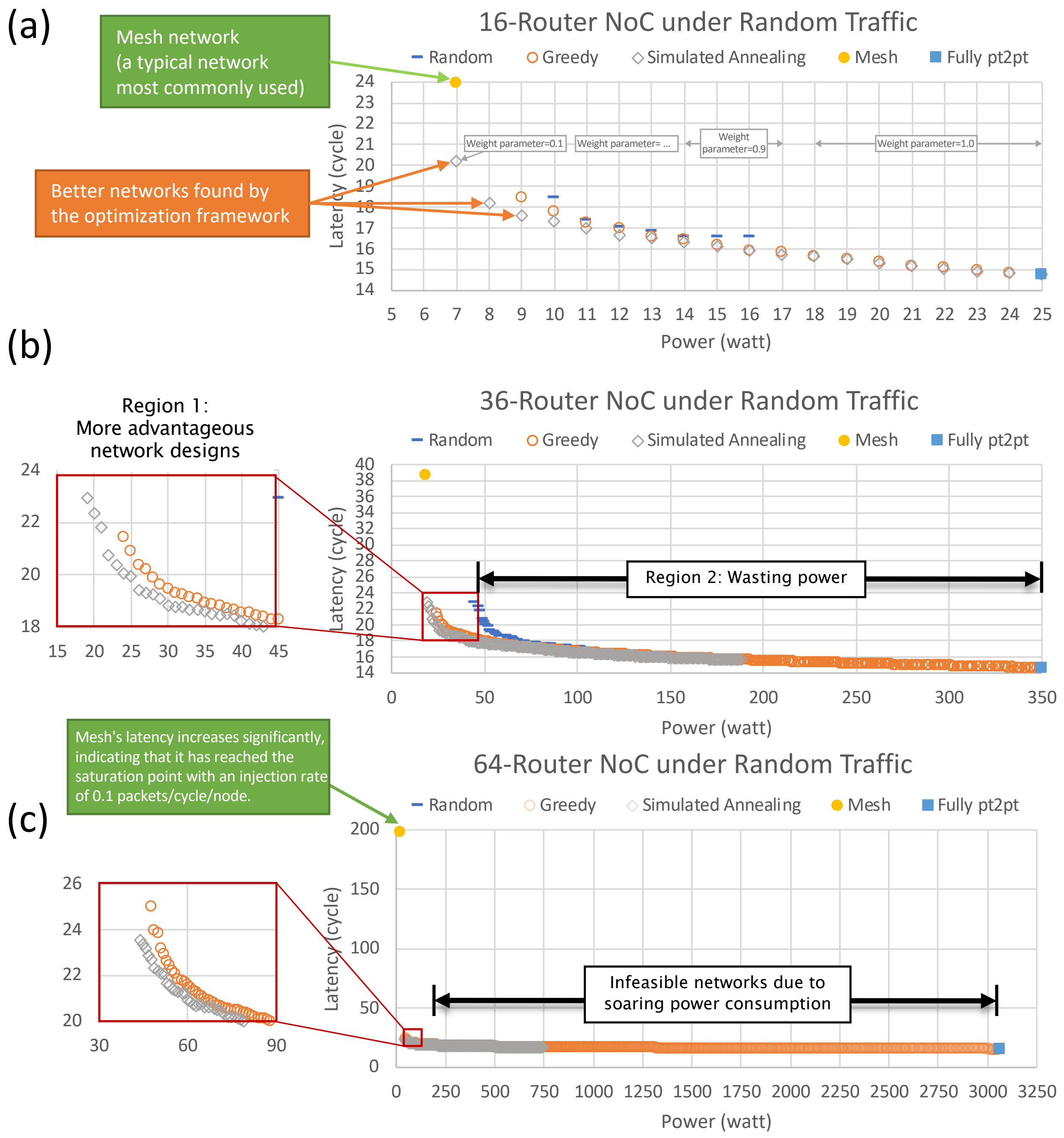

Figure 5. The Pareto-optimal front results for (a) 16-router networks, (b) 36-router networks, and (c) 64-router networks under random traffic. In addition to three optimization algorithms, a mesh network and a fully-connected point-to-point (pt2pt) network are also plotted. The simulated annealing uses different weight parameters (0.1, 0.2, ..., 1.0) and combines the results to form the Pareto-optimal front, e.g., (a) for power 7, weight=0.1, for power 15, weight=0.9, and for power 22, weight=1.0.

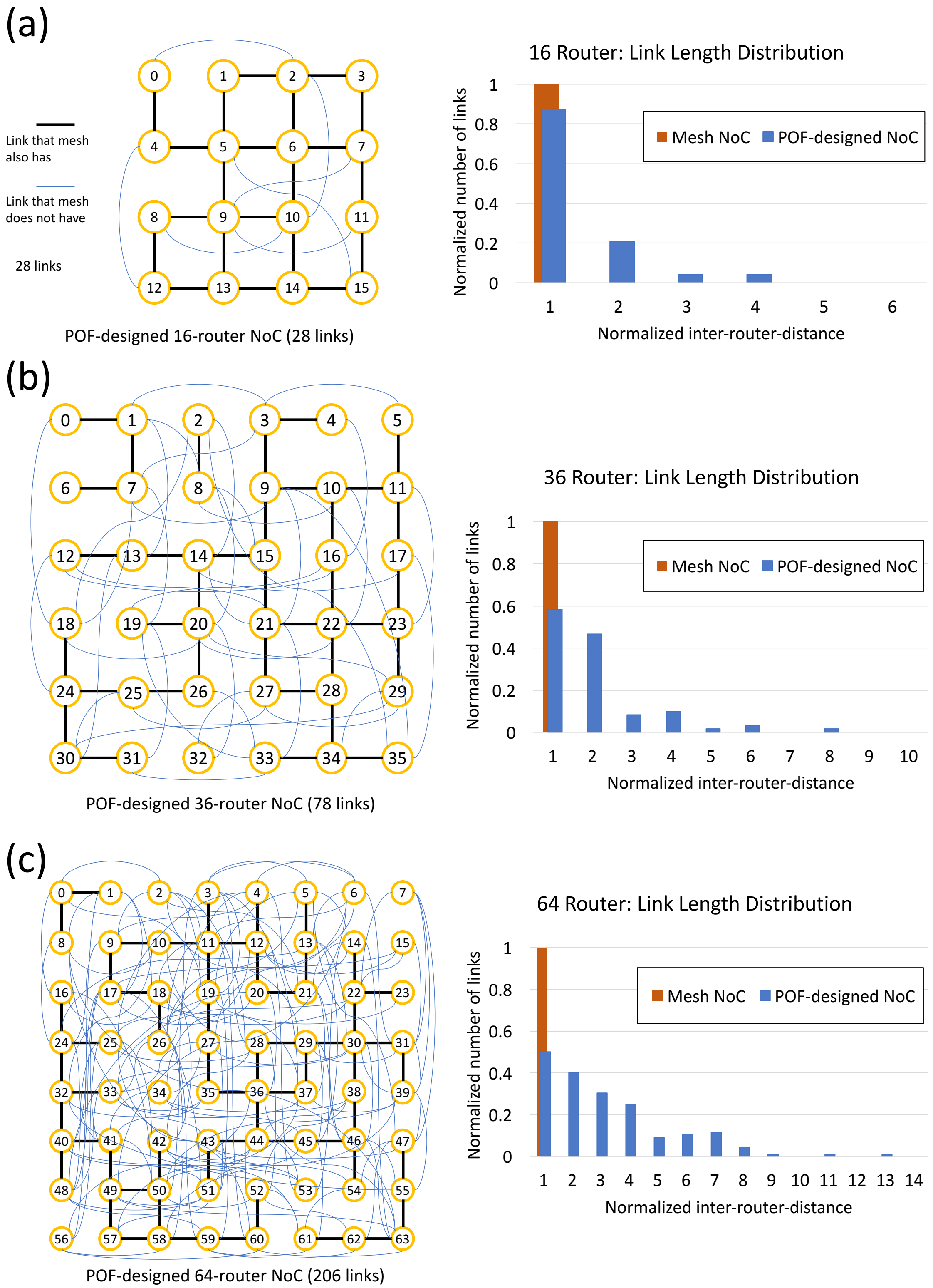

Figure 6. Compared to mesh topology, POF-designed NoCs add more and longer links in place of shorter links as the network size increases (see link length distribution histograms above). The total number of links are: (a) 16 routers: 28 for POF-NoC versus 24 for mesh, (b) 36 routers: 78 for POF-NoC versus 60 for mesh, and (c) 64 routers: 206 for POF-NoC versus 112 for mesh.

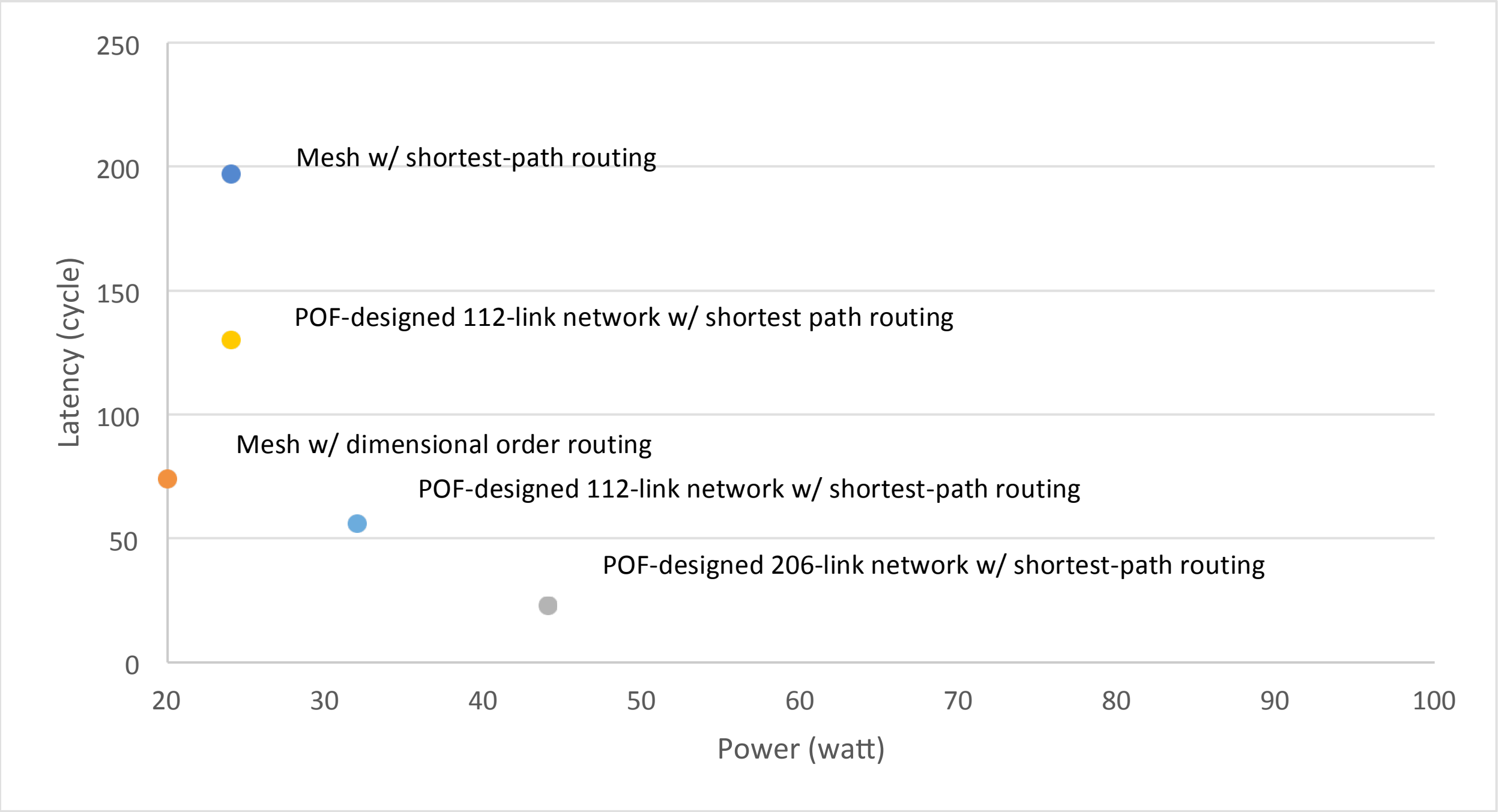

Figure 7. Two POF-designed NoCs with the restriction of 112 links compared with two mesh networks.

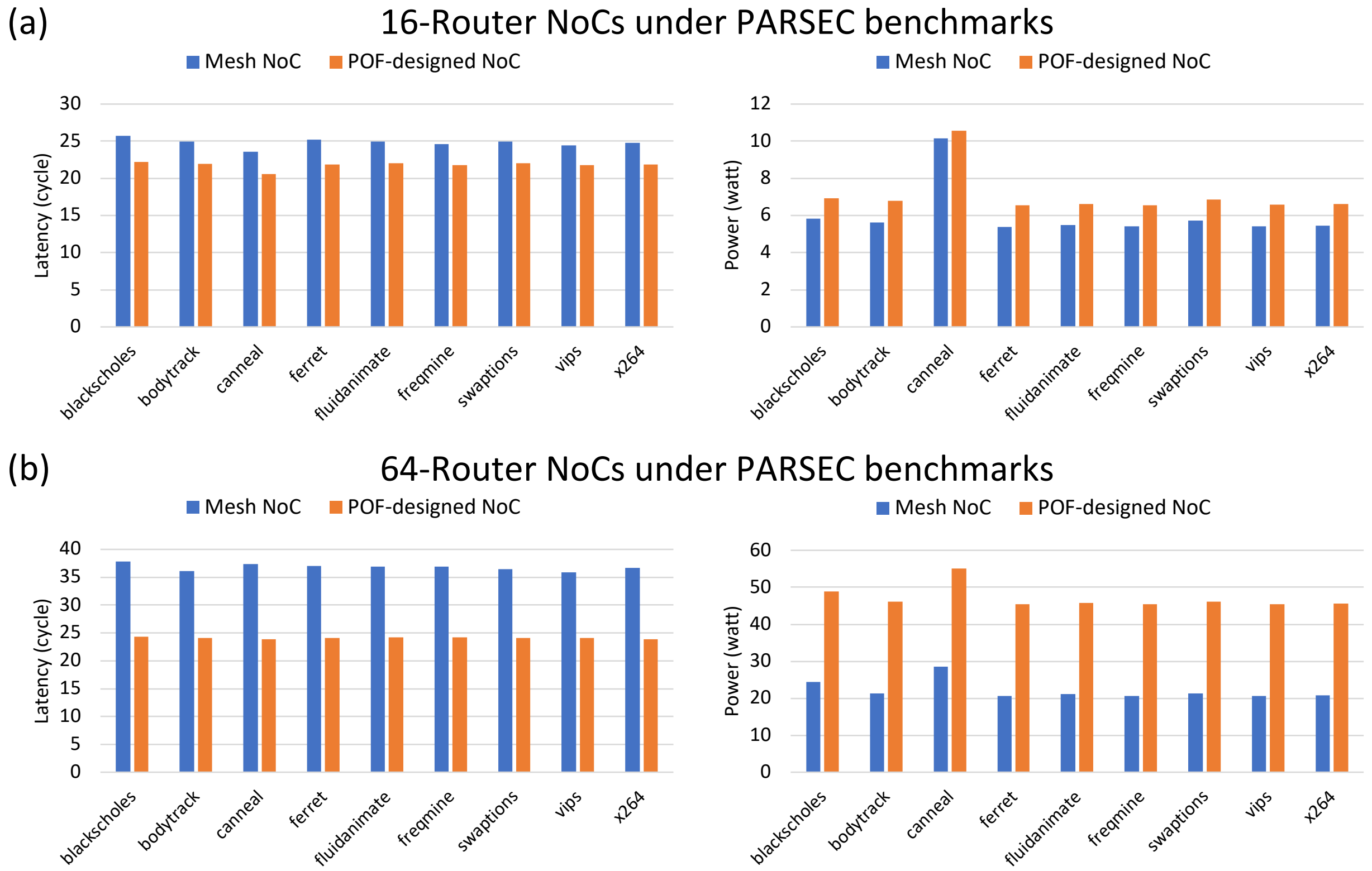

Figure 8. Average network latency and power consumption comparison between mesh NoCs (blue bars) and POF-designed NoCs (orange bars) using the PARSEC benchmark suite [13, 14] running on a full system cycle accurate gem5 simulator [12] and Orion2.0 power simulator [16].